\documentstyle[amstex,amssymb,preprint,aps]{revtex}

\begin{document}

\title{New Upper Limit of Terrestrial Equivalence Principle Test for
Rotating Extended Bodies}
\author{Z. B. Zhou$^{\text{a}}$, J. Luo$^{\text{a *}}$, Q. Yan$^{\text{a}}$,
Z. G. Wu$^{\text{a}}$, Y. Z. Zhang$^{\text{b,c}}$, Y. X. Nie$^{\text{d}}$ }
\address{$^{\text{a}}$Department of Physics, Huazhong Univeristy of Science\\
and Technology, Wuhan 430074, China\\
$^{\text{b}}$CCAST(World Lab.), P. O. Box 8730, Beijing 100080\\
$^{\text{c}}$Institute of Theoretical Physics, Chinese Academy of Sciences,\\
Beijing 100080, China\\
$^{\text{d}}$Institute of Physics, Chinese Academy of Sciences, Beijing\\
100080, China}

\maketitle

\begin{abstract}
Improved terrestrial experiment to test the equivalence principle for
rotating extended bodies is presented, and a new upper limit for the
violation of the equivalence principle is obtained at the level of 1.6$%
\times $10$^{\text{-7}}$, which is limited by the friction of the rotating
gyroscope. It means the spin-gravity interaction between the extended bodies
has not been observed at this level.
\end{abstract}


\section{Introduction}

The equivalence principle (EP), as one of the fundamental hypotheses of
Einstein's general relativity, has been tested by many experiments,
including torsion balance scheme \cite{R1,R2,R3,R4} and free-fall one \cite%
{R5,R6,R7}. Lunar laser ranging from the Earth to the Moon has provided up
to now the most accurate test of the EP to 5$\times $10$^{\text{-13 }}$\cite%
{R8}. Recently, some different tests of EP for gravitational self-energy and
spin-polarized macroscopic objects have been reported \cite{R9,R10,R11}.
However, in all of the experiments including the Satellite Test of the
Equivalence Principle (STEP) and the Galileo Galilei (GG) space projects as
well as the MICROSCOPE space mission \cite{R12,R13,R14}, the test masses are
all non-rotating.

It is well known that spin-interactions of elementary particles, spin-orbit
coupling and spin-spin coupling, have been studied in both theory and
experiment. Furthermore, the spin-gravitational couplings, i.e. the
spin-gravitoelectric coupling and the spin-gravitomagnetic coupling, and the
spin-rotation coupling between intrinsic spins have been also investigated
for a long time \cite{R15,R16,R17,R18,R19,R20}.

Over the last few years there has been a growing interest in effects of
gravitational fields on gyroscopes \cite{R21,R22,R23,R24}. From the
experimental point of view, the NASA/Stanford Relativity Mission Gravity
Probe B (GP-B) experiment will provide two extremely precise tests of
general relativity based on observations of four identical superconducting
gyroscopes in a satellite in a 400 miles polar orbit around the Earth \cite%
{R25}. These gyroscopes are carefully isolated from nearly all sources of
Newtonian torques, and their residual drift is presented less than 0.020
marc-second/year for a gyroscope in a fully inertial orbit \cite{R26}.
General relativity predicts that, though isolated from external torques, the
spin axes of these gyroscopes will precess with respect to a distant
inertial reference frame at a rate of 6.6 arc-second/year for the geodetic
effect, and 0.042 arc-second/year due to frame dragging. Cerdonio et al.
have proposed a novel detector (gyromagnetic electron gyroscope) to locally
detect the frame dragging due to the terrestrial rotation \cite{R27}.
Recently, Zhang et al. also developed a phenomenological model for the
spin-spin interaction between rotating extended bodies, which predicts the
effect of the spin-spin coupling on the orbital acceleration of the
gyroscope free-falling in gravitational field rather than the spin
procession of the gyroscope\cite{R28}. In the mode, a dimensionless
parameter representing the strength of violation of EP can be defined as
follows
\begin{equation}
\eta _{\text{s}}=\frac{\Delta g}{g}=\kappa \left( \frac{\stackrel{%
\rightharpoonup }{S}_{\text{1}}\cdot \stackrel{\rightharpoonup }{S}_{\text{e}%
}}{Gm_{\text{1}}M_{\text{e}}R_{\text{1}}}-\frac{\stackrel{\rightharpoonup }{S%
}_{\text{2}}\cdot \stackrel{\rightharpoonup }{S}_{\text{e}}}{Gm_{\text{2}}M_{%
\text{e}}R_{\text{2}}}\right) \text{ ,}  \label{Eq.1}
\end{equation}%
where {\it G} is the Newtonian gravitational constant, {\it m}$_{\text{1}}$,
{\it m}$_{\text{2}}$ and {\it M}$_{\text{e}}$ are the masses of the two
gyroscopes and the Earth, respectively, and $\stackrel{\rightharpoonup }{S}_{%
\text{1}}$, $\stackrel{\rightharpoonup }{S}_{\text{2}}$, and $\stackrel{%
\rightharpoonup }{S}_{\text{e}}$ are their spin angular momentums, {\it R}$_{%
\text{1}}$ and {\it R}$_{\text{2}}$ are the distances between the centers of
the two gyroscopes and the Earth, respectively, and the parameter $\kappa $
represents the universal coupling factor for the spin-spin interaction for
rotating extended bodies. As pointed out in Refs. 28 and 29, the
phenomenological model developed by Zhang et al. is to investigate the
effect of the spin-spin coupling on the orbital acceleration of the rotating
gyroscope free falling in gravitational field, which is different from the
aim of the GP-B.

A preliminary double free-fall (DFF) experiment to test the EP for rotating
extended bodies, in which two gyroscopes with differing rotating senses drop
freely, has been performed, and the results show that the EP is still valid
for rotating extended bodies at the level of 2$\times $10$^{\text{-6}}$ \cite%
{R29}. A main limit of preliminary experimental precision has been
proved to come from the pump outgassing effect due to the
asymmetrical outgassing for the two tubes. In the initial
experimental setup, the vacuum pump system is set in the top part
of the tube. In this case, when the test masses fall through the
tee part of the tube ($\backsim $ 0.3 s free-fall), a force due to
the pump outgassing will deflect the test masses or lead them to a
more complex motion \cite{R30}. An abrupt acceleration change of
about 20 mGal is observed at this height. To avoid it, the pump
system is moved down to the bottom part of the tube, and meantime
the vacuum level is improved from initial 50 mPa to 2 mPa. At the
same time, the vibration excited by the operating pump is
effectively isolated by a rubber-gas-steel isolator, and the
isolation ratio is measured about -30 dB to -60 dB in the range of
above 3 Hz. And then, the effective falling height is prolonged
from initial 20 cm (0.2 s) to 9 m (1.0 s). In this article, error
sources of our DFF experiment will be carefully discussed and a
new upper limit of the EP for rotating extended bodies will be
presented.

\section{Experimental Description and Error Analysis}

A Michelson-type interferometer including a frequency-stabilized
He-Ne laser beam with a relative length standard of 1.3$\times
$10$^{\text{-8}}$ is used to monitor the differential vertical
displacement between two gyroscopes, in which one is rotating and
another is non-rotating, and then the interference fringes are
sampled by means of a 10 MHz 12-bit AD card combined with an
external rubidium atomic clock, and then stored in a computer. The
diameter of the laser beam is collimated about 3.0 mm so that the
beam wavefront effect can be neglected. An aligned verticality is
kept within 50$^{\prime \prime }$ for each laser beam, and the
maximum uncertain differential acceleration due to the aligned
verticality is below 20 ${\sc \mu }$Gal.

Each of the two test masses consists of a steel gyroscope with a mass of
420.0$\pm $2.5 g, a diameter of about 55 mm and a height of about 32 mm,
which together with a corner-cube-retroreflector (CCR) of 76.4$\pm $0.4 g is
sealed in an aluminum frame of 159.4$\pm $0.9 g. Tinned copper wires with a
diameter of 0.25 mm are used to suspend the test masses, and the initial
suspending differential height between them is kept within 1 mm, which
implies the vertical gravity gradient correction is about 0.3 ${\sc \mu }$%
Gal. The test mass with a non-rotation rotor is released about 3 ms before
the rotating one, which sets a systematic error of about 0.3 $\mu $Gal due
to the finite speed of light. The rotating gyroscope is driven by a DC motor
and its rotating speed is kept at (17000$\pm $200) rpm.

An uncertain acceleration due to the residual gas drag effect is less than
0.2 ${\sc \mu }$Gal at $p$ = 2 mPa and $T$ = 300 K \cite{R29}. In addition,
the outgassing effect on the dropped objects should be carefully considered
because of the continuous operation of a turbo-molecular pump with a full
rated pumping speed $v_{\text{p}}$ of 1500 L/s. The acceleration
contribution for a single dropped object can be estimated as follows \cite%
{R31}
\begin{equation}
a\leq R\rho v_{\text{p}}V/m  \label{Eq.2}
\end{equation}%
where $R$ represents the ratio of the surface of the dropped object and the
inner surface of the vacuum tube, and is about 0.04 here, $\rho $ and $V$
are the residual gas density and mean gas particle speed, respectively.
Then, the acceleration for a single dropped object is about 100 $\mu $Gal at
$p$ = 2 mPa. Fortunately, the DFF scheme can reduce the common mode effect,
and a differential acceleration for the two test masses due to the
outgassing effect depends on both the outgassing difference and the gas
density difference between the two tubes. It is very difficult to calculate
the real difference due to the complex flux motion exactly. However, it can
be roughly estimated based on the pressure difference between the two tubes.
When the turbo-molecular pump runs normally, the pressure of the bottom part
of the tube (close to the pump) is measured about 1.7 mPa, and at the same
time, the top parts of the two tubes are measured about 8.7 mPa and 5.6 mPa,
respectively. This means that the outgassing speeds in the two tubes are
about 0.61 $v_{\text{p}}$ and 0.39 $v_{\text{p}}$ if the pressure
distribution in both tubes is the same. So in this assumption, the
acceleration difference for the two test masses due to the outgassing effect
is estimated less than 22 $\mu $Gal.

A possible lifting force for a rotating rotor due to the residual gas flow's
circulation can be calculated based on the Zhukovskii's theorem, and this
effect can be neglected here \cite{R29}. A possible horizontal velocity
difference $\Delta v_{\text{h}}$ is estimated smaller than 4.2 mm/s
according to the change of the interference pattern intensity during
free-fall of test masses, and an acceleration difference due to the
Coriolis's effect is less than 54 $\mu $Gal \cite{R29}. It means that the
horizontal velocity difference would have to be monitored in the further
experiment with a higher precision.

The silent amplitude spectrum of the seismic noise in our laboratory
contributes an uncertainty of about 1 ${\sc \mu }$Gal to the final
experiment result \cite{R32}. But the mechanical vibration modes of the
optical measurement system are dominant due to excitation of the vacuum
pumps. Figure 1 is a typical residual differential displacement of the DFF
experiment between both non-rotating test masses, in which the linear term
has been subtracted. Three main modes have been observed about 16.8 Hz, 36.6
Hz, and 96.1 Hz, and their amplitudes comes to about 0.05 $\mu $m, which
contribute an uncertain acceleration of 8 $\mu $Gal based on the following
equation \cite{R30}%
\begin{equation}
\Delta g_{\text{f}}=\frac{120x_{\text{n}}}{\omega _{\text{n}}T^{\text{3}}}%
\sqrt{1+\frac{12}{\omega _{\text{n}}^{2}T^{\text{2}}}+(\frac{12}{\omega _{%
\text{n}}^{2}T^{\text{2}}})^{\text{2}}}\cos (\frac{\omega _{\text{n}}T}{2}%
+\phi _{\text{n}})\cos (\frac{\omega _{\text{n}}T}{2}+\text{tg}^{-1}\frac{%
12-\omega _{\text{n}}^{2}T^{\text{2}}}{6\omega _{\text{n}}T}),
\end{equation}%
where $x_{\text{n}}$, $\omega _{\text{n}}$, and $\phi _{\text{n}}$ represent
the amplitude, the angular frequency, and the phase of the high-frequency
vibration, respectively, and $T$ is the effective time length. Parabolic
curve fitting result shows that the differential acceleration $\Delta g_{%
\text{N-N}}$ is -58 $\mu $Gal, which is consistent with the total systematic
uncertainty, as listed in Table 1, of 64 $\mu $Gal.

\section{Experimental Results}

Figure 2(a) and 2(b) are, respectively, typical residual differential
displacements of the DFF experiment with non- and left-rotating (spin vector
pointing upward) gyroscopes, and non- and right-rotating (spin vector
pointing downward) ones. From both figures, it is very clear that there are
a dominant slow frequency motion but not a parabolic term. This slow
frequency (1.6$\pm $0.2 Hz) motion is confirmed to be resulted from the
friction coupling between the rotating rotor and the aluminum frame by
observing the motion of the reflected beam. Since effective free-fall
duration in our DFF experiment is about 1 s and the correlative coefficient
between a parabolic term ($\Delta g$ $\backsim $ 100 $\mu $Gal) and a
harmonic term of 1.6 Hz ($x_{\text{n}}$ $\backsim $ 0.06 $\mu $m) comes to
0.3 $\backsim $ 0.4, the slow frequency fluctuation is very difficult to be
subtracted by fitting. Nevertheless, a maximum uncertain acceleration due to
the slow motion can be estimated based on Eq. (3), and its effect comes to
about 130 $\mu $Gal ($x_{\text{n}}\backsim $ 0.06 $\mu $m) for the non- and
left-rotating gyroscopes and 150 $\mu $Gal ($x_{\text{n}}\backsim $ 0.07 $%
\mu $m) for the non- and right-rotating gyroscopes, respectively. The
fitting results of the both residual curves show that the differential
acceleration between the non- and left-rotating gyroscopes $\Delta g_{\text{%
N-L}}$ is 0.48 $\mu $Gal, and that between the non- and right-rotating
gyroscopes $\Delta g_{\text{N-R}}$ is -110 $\mu $Gal, which are consisted
with the systematic uncertainty of 160 $\mu $Gal as listed in Table 1. In
addition, a high-frequency mechanical vibration of the CCR at the frequency
of the rotating rotor, caused by the friction coupling, has also been
observed and its modulation amplitude is in the order of 0.1 $\mu $m, which
contributes an uncertainty of about 5 $\mu $Gal. Fortunately, the limit can
be suppressed by a factor of sin($\omega _{\text{n}}t_{\text{s}}/2$)/($%
\omega _{\text{n}}t_{\text{s}}/2$) by means of a time-domain
data-smoothing-process, here $t_{\text{s}}$ is the time length of the
smoothing-process.

Based on above statement, the EP is still valid at the level of 1.6$\times $%
10$^{\text{-7}}$ for rotating extended bodies, which is improved by over one
order for preliminary experiment result \cite{R23}, and the spin-spin
interaction between the rotating extended bodies has not been observed at
this level. According to the Eq. (1) and the approximately uniform sphere
mode of the Earth, it can be concluded that the coupling factor $\kappa \leq
$1.6$\times $10$^{\text{-19 }}$kg$^{\text{-1}}$, which sets a new upper
limit for the spin-spin interaction between a rotating extended body and the
Earth.

\section{Discussions}

Our experiment precision is mainly limited by the mechanical friction of the
gyroscope, which could be improved by choosing better gyroscope or extending
the free-fall duration. As pointed out in the introduction section, GP-B
mission measures the spin axes precession of spinning fused-quartz
gyroscopes with respect to an inertial reference frame, partly due to the
geodetic effect, and partly due to frame dragging. In general relativity the
both effects have small, non-zero values even without violation of the EP.
However, based on the model developed by Zhang et al., a perigean precession
of the gyroscope in the GP-B experiment is not greater than 100
arc-second/year based on current experimental result $\kappa \leq $1.6$%
\times $10$^{\text{-19 }}$kg$^{\text{-1}}$. Nevertheless, the GP-B
experiment measures the spin axis precession rather than the orbital motion
of the gyroscope. As proposed in Ref. 28, a possible scheme is to put a
non-spinning shell surrounding a spinning gyroscope in a satellite, and the
motion of the spinning gyroscope with respect to the non-spinning reference
frame could be monitored using a SQUID or an inertial sensor \cite{R33,R34}.
If the gap between the gyroscope and the reference shell can be measured in
the level of 1 nm/year, and the coupling factor $\kappa $ between the
spin-spin interaction between the rotating extended bodies could be tested
in the level of 10$^{\text{-31}}$ kg$^{\text{-1}}$, which is improved by
about 12 orders.

{\bf Acknowledgments} We are grateful to Prof. W. R. Hu and A.
Ruediger for their discussion and useful suggestion. This work was
supported by the Ministry of Science and Technology of China under
Grant No: 95-Yu-34 and 19990754 and by the National Natural
Science Foundation of China under Grant No: 19835040, 10175070,
and 10047004.

\begin{figure}[tbp]
\caption{Residual differential displacement of the double free-fall
experiment between both non-rotating test masses. The fluctuation is due to
the mechanical vibration modes of optical measurement system. The parabolic
curve fitting shows that the differential acceleration is about 58 $\protect%
\mu $Gal.}
\label{FIG 1}
\end{figure}

\begin{figure}[tbp]
\caption{(a) Residual differential displacement of the double free-fall
between non- and left-rotating gyroscopes. The parabolic curve fitting shows
that the differential acceleration is about 0.48 $\protect\mu $Gal. (b)
Residual differential displacement between non- and right-rotating
gyroscopes. The parabolic curve fitting shows that the differential
acceleration is about -110 $\protect\mu $Gal.}
\label{FIG 2}
\end{figure}

\begin{table}[tbp]
\caption{Summary of systematic errors of the equivalence principle test for
rotating extended bodies.}
\label{TABLE 1}
\end{table}

\begin{tabular}[b]{cc}
\hline\hline
Systematic Error & \ Uncertainty\ ($\mu $Gal) \\ \hline
\multicolumn{1}{l}{Length standard of laser} & $\backsim $ 13 \\
\multicolumn{1}{l}{Verticality of laser beam} & $\leq $ 20 \\
\multicolumn{1}{l}{Outgassing effect} & $\backsim $ 22 \\
\multicolumn{1}{l}{Horizontal motion} & $\leq $ 54 \\
\multicolumn{1}{l}{Mechanical vibrations} & $\backsim $ 8 \\
\multicolumn{1}{l}{Total for non-rotating} & $\leq $ 64 \\ \hline
\multicolumn{1}{l}{Friction coupling} & $\leq $ 150 \\
\multicolumn{1}{l}{Total for rotating} & $\leq $ 160 \\ \hline\hline
\end{tabular}


\begin{references}
\bibitem{R1} R. V. E\"{o}tv\"{o}s, D. Pekar, and E. Fekete, Ann. Phys.
(Leipzig) 68 (1922) 11.

\bibitem{R2} P. G. Roll, R. Krotkov, and R. H. Dicke, Ann. Phys. (N.Y.) 26
(1964) 442.

\bibitem{R3} V. B. Braginsky and V. I. Panov, Sov. Phys. JETP 34 (1972) 463.

\bibitem{R4} Y. Su et al., Phys. Rev. D 50 (1994) 3614.

\bibitem{R5} T. M. Niebauer, M. P. McHugh, and J. E. Faller, Phys. Rev.
Lett. 59 (1987) 609.

\bibitem{R6} K. Kuroda and N. Mio, Phys. Rev. Lett. 62 (1989) 1941.

\bibitem{R7} S. Carusotto et al., Phys. Rev. Lett. 69 (1992) 1722.

\bibitem{R8} J. O. Dickey et al., Science 265 (1994) 482.

\bibitem{R9} S. Baessler et al., Phys. Rev. Lett. 83 (1999) 3585.

\bibitem{R10} R. C. Ritter et al., Phys. Rev. D 42 (1990) 977.

\bibitem{R11} W. D. Ni et al., Phys. Rev. Lett. 82 (1999) 2439; L. S. Hou
and W. T. Ni, Mod. Phys. Lett. A 16 (2001) 763.

\bibitem{R12} STEP : Testing the Equivalence Principle in Space,
Proceedings, edited by R. Reinhard (Pisa, 1993).

\bibitem{R13} A. M. Nobili et al., J. Astronaut. Sci. 43 (1995) 219.

\bibitem{R14} P. Touboul and M. Rodrigues, Class. Quantum Grav. 18 (2000)
2487.

\bibitem{R15} I. Yu Kobzarev and L. B. Okun, JETP 16 (1963) 1343.

\bibitem{R16} D. J. Wineland and N. F. Ramsey, Phys. Rev. A 5 (1972) 821; C.
J. Berglund et al., Phys. Rev. Lett. 75 (1995) 1879; A. N. Youdin et al.,
Phys. Rev. Lett. 77 (1996) 2170;

\bibitem{R17} C. G. de Oliveira and J. Tiomno, Nuovo Cimento 24 (1962) 672;
B. M. Barker and R. F. O'Connell, Phys. Rev. D 12 (1975) 329; L. H. Ryder,
Gen. Rel. Grav. 31 (1999) 775.

\bibitem{R18} B. Mashhoon, Nature 250 (1974) 316; Phys. Rev. D 10 (1974)
1059; Phys. Rev. D 11 (1975) 2679.

\bibitem{R19} S. A. Werner, J. L. Staudenmann and R. Colella, Phys. Rev.
Lett. 42 (1979) 1103; T. L. Gustavson, P. Bouyer and M. A. Kasevich, Phys.
Rev. Lett. 78 (1997) 2046.

\bibitem{R20} B. Mashhoon, Phys. Lett. A 198 (1995) 9; B. Mashhoon, R.
Neutze, M. Hannam and G. E. Stedman, Phys. Lett. A 249 (1998) 161; B.
Mashhoon, Phys. Rev. A 47 (1993) 4498.

\bibitem{R21} K. D. Krori, T. Chaudhury, and C. R. Mahanta, Phys. Rev. D 42
(1990) 3584.

\bibitem{R22} B. Mashhoon, Class. Quantum Grav. 17 (2000) 2399; B. Mashhoon,
Gen. Rel. Grav. 31 (1999) 681.

\bibitem{R23} L. Herrera, F. M. Paiva, and N. O. Santos, Class. Quantum
Grav. 17 (2000) 1549.

\bibitem{R24} F. Sorge, D. Bini, and F. de Felice, Class. Quantum Grav. 18
(2001) 2945.

\bibitem{R25} S. Buchman et al., Adv. Space Res. 25 (2000) 1177.

\bibitem{R26} S. Buchman et al., Phys. B 280 (2000) 497.

\bibitem{R27} M. Cerdonio, G. A. Prodi, and S. Vitale, Gen. Rel. Grav. 20
(1988) 83.

\bibitem{R28} Y. Z. Zhang, J. Luo, and Y. X. Nie, Mod. Phys. Lett. A 16
(2001) 789.

\bibitem{R29} J. Luo, Y. X. Nie, Y. Z. Zhang, and Z. B. Zhou, Phys. Rev. D
65 (2002) 042005.

\bibitem{R30} Z. B. Zhou, PhD thesis, Huazhong University of Science and
Technology (in Chinese), 2001.

\bibitem{R31} T. M. Niebauer et al., Metrologia 32 (1995) 159.

\bibitem{R32} Z. B. Zhou et al., Chin. Phys. Lett. 18 (2001) 10.

\bibitem{R33} V. Josselin, M. Rodrigues, and P. Touboul, Acta Astronutica 49
(2001) 95.

\bibitem{R34} A. Cavalleri et al., Class. Quantum Grav. 18 (2001) 4133.
\end{references}
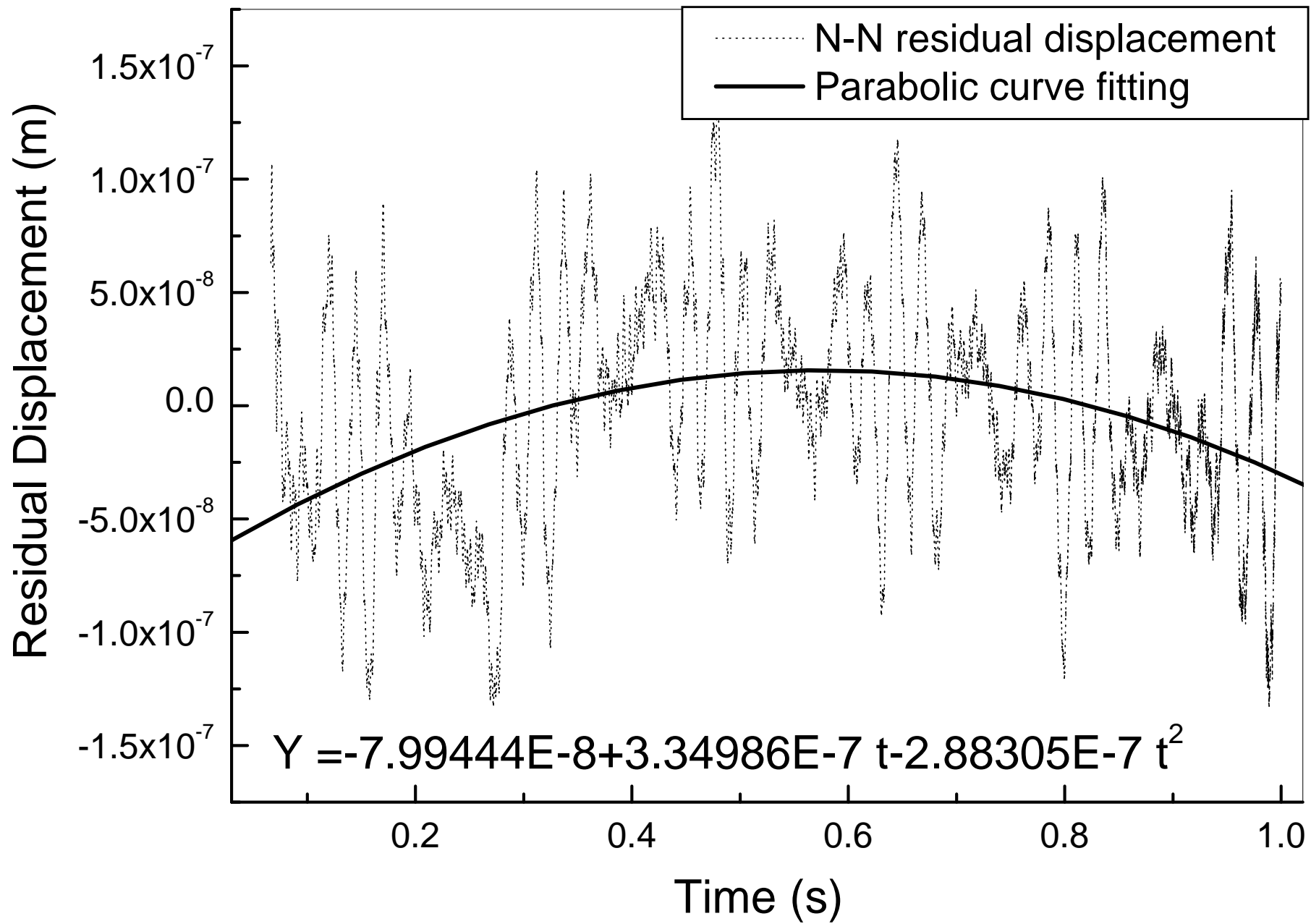
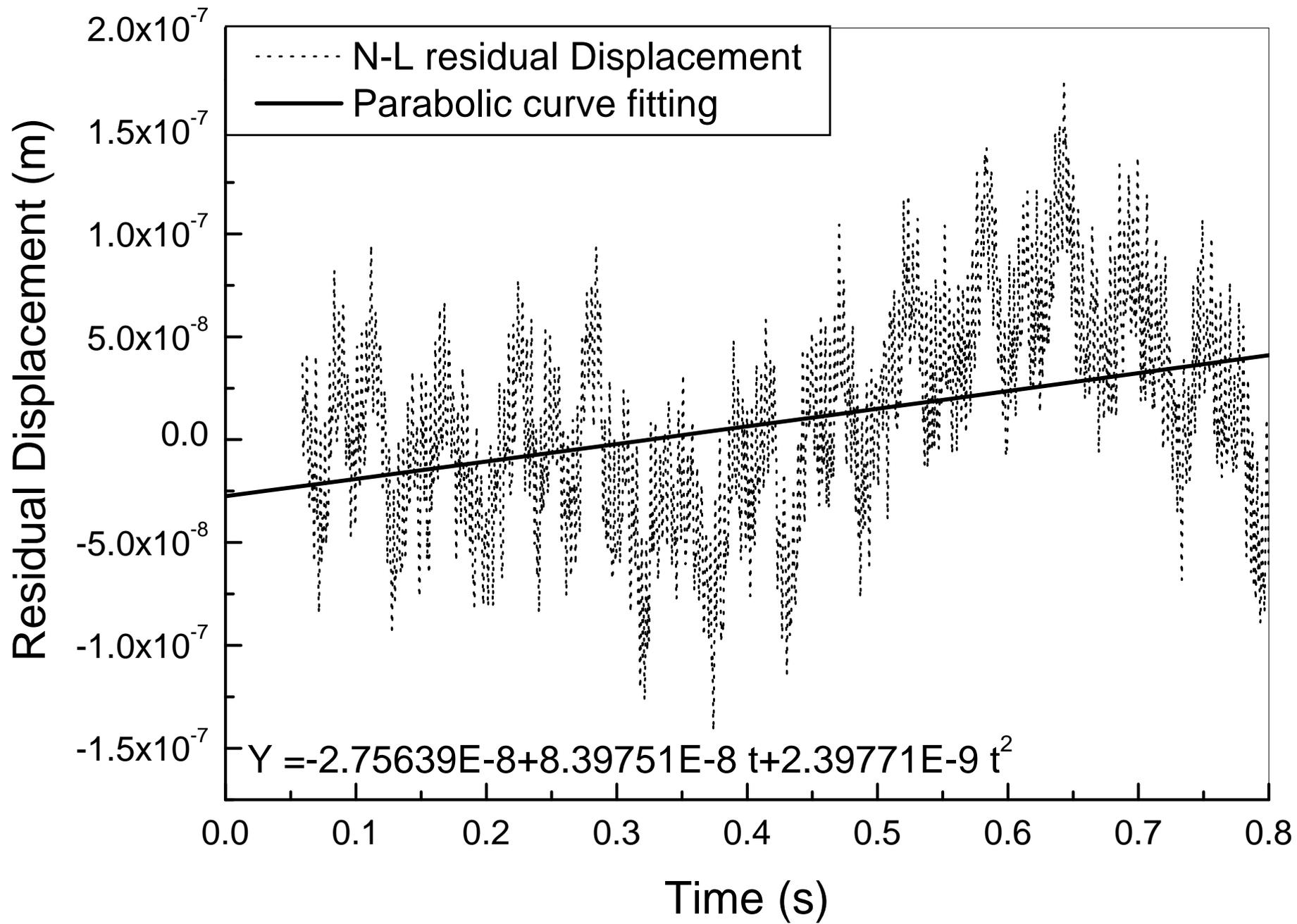
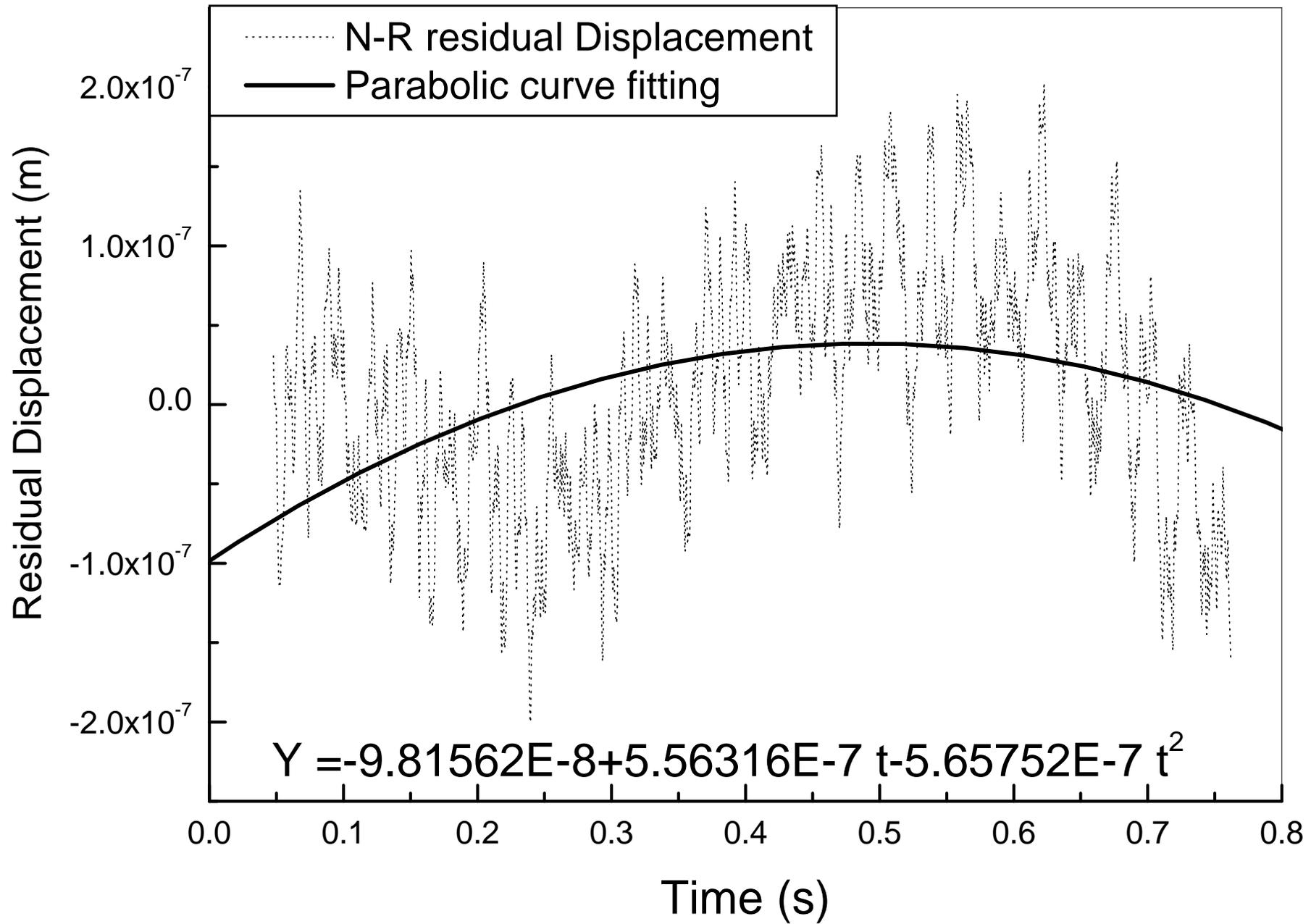
\end{document}